\documentclass[
twocolumn,
showkeys,
reprint,
superscriptaddress,
amsmath,amssymb,
aps, pre
]{revtex4-2}

\usepackage{graphicx}
\usepackage{dcolumn}
\usepackage{bm}
\usepackage{hyperref}
\usepackage{standalone}
\usepackage[mathlines]{lineno}

\begin{document}

\title{Coevolution of epidemic dynamics and network topology driven by disease fatality and waning immunity}

\author{ThankGod I. S. Ikpe}
\affiliation{Graduate School of Information Sciences, Tohoku University, Sendai, 980-8579, Japan}
\author{Takayuki Hiraoka}
\affiliation{Department of Computer Science, Aalto University, Espoo, 00076, Finland}
\author{Naoya Fujiwara}
\affiliation{Graduate School of Information Sciences, Tohoku University, Sendai, 980-8579, Japan}
\affiliation{PRESTO, Japan Science and Technology Agency, Kawaguchi, 332-0012, Japan}

\date{\today}

\date{\today}

\begin{abstract}
Epidemics on complex networks have been shown to exhibit dynamics that are strongly influenced by the topology of the network. However, it remains unclear how the topology of the network is influenced by epidemic properties, such as waning immunity and disease fatality, coupled with demographic changes. To explore the interplay between epidemic dynamics and network topology, we develop an agent-based model of a fatal infectious disease with an imperfect immunity on an initially scale-free network that evolves via demographic and epidemic-induced changes. We show that this model undergoes an epidemic transition between a phase in which the disease eventually dies out and a phase in which it becomes endemic. Moreover, the network may lose its scale-free property as a result of the epidemic spreading, giving rise to a topological transition from a power-law to a non-power-law degree distribution. We validate our findings using heterogeneous mean-field approximations of the agent-based model. These results highlight the importance of accounting for network evolution in epidemic models and advance our understanding of coevolving dynamical systems, in which disease dynamics and network topology continuously shape one another.
\end{abstract}

\maketitle

\section{Introduction}

Network epidemiology---the study of epidemics on complex networks---gained traction following the discovery of scale-free networks~\cite{Barabasi-Albert, Barabasi-Jeong-Albert}. Since this discovery provided a framework for understanding the structure of real-world networks such as social contact networks and transportation networks~\cite{liljeros2001sexual, mislove2007measurement, guimera2005worldwide}, the spread of epidemics on such complex networks has been widely studied~\cite{PastorSatoras-Vespignani-1, PastorSatoras-Vespignani-2, Newman2002, Keeling-Eames2005, PastorSatoras-et.al-ReviewPaper}. Traditional compartmental models assume that all individuals in a population have equal opportunity of making contact with others~\cite{Kermack-McKendrick1927, AndersonMay1991, Diekmann1990,Hethcote}; in contrast, scale-free networks are characterized by heavy-tailed degree distributions, meaning that a few highly connected nodes, i.e., hubs, play a disproportionate role in disease transmission. Scale-free networks capture the heterogeneity of real-world populations, thereby overcoming the limitations of traditional approaches~\cite{WattsStrogatz1998, Keeling-Eames2005, Bansal2007}.

While epidemics have been studied predominantly assuming a closed population, real-world populations are subject to demographic changes. Demographic processes such as births, deaths, immigration, and emigration continually add and remove individuals from the population. The effects of these demographic factors on the dynamics of epidemics have been extensively studied in traditional approaches using compartmental models based on differential equations~\cite{hethcote1973asymptotic, rock2014dynamics, Xie-Ishfaq-ThankGod-Elza-Seno}. In particular, Xie et al.~\cite{Xie-Ishfaq-ThankGod-Elza-Seno} showed that, under certain conditions, these demographic changes can either suppress or exacerbate the endemicity of epidemics.

In network epidemic models, demographic processes can be embodied by network evolution rules, i.e., the ways in which nodes are added to and/or removed from the network. Previous studies have addressed how the addition and removal of nodes lead to global topological reorganization~\cite{DJD_Price1965, Barabasi-Albert, Dorogovtsev-et.al, Budnick-el.al_Isreal2022, Moore-Ghoshal-Newman, Bauke-Moore-et.al2011, Lee.M.J-et.al2022}. For example, Lee et al.~\cite{Lee.M.J-et.al2022} reported a deviation from the initial power-law degree distribution when hubs were targeted for removal, while the scale-free structure is preserved under a non-targeted removal process, further emphasizing the importance of hubs in network resilience. Note that the above studies focused solely on network evolution and did not consider epidemics.

Several previous works have addressed the interplay between epidemic dynamics and demographic processes in network models. Refs.~\cite{LI-Chi-Yang2014, HUANG-Chen-Chen2017,Hu-et.al} have incorporated births and natural deaths into epidemic models that assume waning immunity. In these works, the authors did not consider deaths from infection; they also assumed that births and deaths would balance out to keep the total population constant, and that the degree distribution would remain unchanged over time. Demirel et al.~\cite{Demirel-Barter-Thilo} considered the dynamics of epidemics on a growing adaptive network. They showed that no epidemic threshold exists in the thermodynamic limit, and that an exponential degree distribution emerges as a result of the feedback between network growth and epidemic spreading mechanisms. In their work, infected nodes either die or remain infected, but never recover. 

No previous work has considered the impact of the interplay between epidemic dynamics---with waning immunity and disease fatality---and the topological evolution of scale-free networks. This feedback represents a bidirectional coupled system where the spread of a disease (dynamics) changes the structure of the population (topology), which in turn alters the future trajectory of the epidemic. This interplay could be important in understanding how diseases spread under adaptive responses, which static network models fail to capture.

In this study, we develop an agent-based model of a fatal infectious disease with imperfect immunity. The interaction between individuals is represented by a network that evolves through the preferential attachment of new nodes via birth, and the removal of nodes that die either naturally or as a result of infection. By combining agent-based simulations and heterogeneous mean-field approximations, we analyze how epidemic processes influence the topological evolution of a scale-free network, and how the evolving network structure subsequently affects disease transmission, persistence, and extinction. Specifically, we focus on the role of waning immunity and infection-induced mortality in shaping this feedback between epidemic dynamics and network topology.

The remaining sections of this paper are organized as follows: In Sec. II, we describe the agent-based model that couples epidemic spreading and network evolution. In Sec. III, we present the results of our simulations. In Sec. IV, we derive a heterogeneous mean-field approximation of the agent-based model, present its numerical solution, and evaluate how accurately the resulting ODE framework reproduces the dynamics observed in the agent-based simulations. Discussions on our findings, the significance of our results, and possible future work are covered in Sec. V.

\section{The Agent Based Model}

\begin{figure}[t]
\centering
\includegraphics[width=0.45\textwidth]{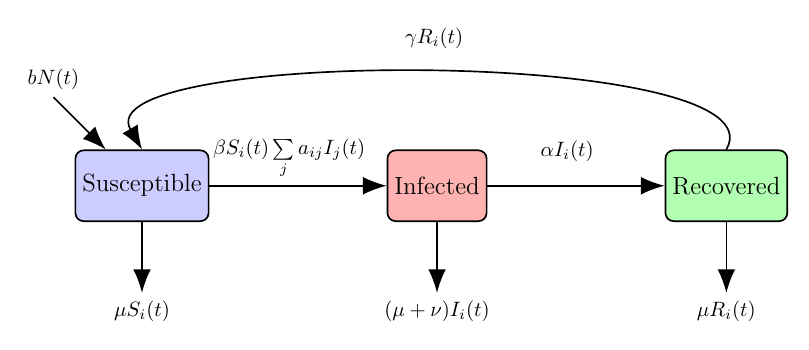}

\caption{Schematic representation of the network-based SIRS epidemic model. Nodes occupy one of three epidemiological states: Susceptible (S), Infected (I), or Recovered (R). Susceptible nodes become infected at rate $\beta$ through contact with infected neighbors. Infected nodes recover with immunity at rate $\alpha$, and recovered nodes lose immunity at rate $\gamma$, returning to the susceptible class. Nodes are introduced through births at rate $b$ and are removed via natural death at rate $\mu$ or disease-induced death at rate $\nu$. The total population at time $t$ is denoted by $N(t)$. The state variables $S_i(t)$, $I_i(t)$, $R_i(t)$ $\in \{0,1\}$ are binary and indicate the epidemiological state of node $i$ at time $t$. The number of infected neighbors of node $i$ is given by $\sum_j a_{ij} I_j(t)$, where $a_{ij}$ is the $(i,j)$-th entry of the adjacency matrix, taking a value of $1$ if nodes $i$ and $j$ are connected and $0$ otherwise.}
\label{fig:schematic}
\end{figure}

We adopt an agent-based model of susceptible-infected-recovered-susceptible (SIRS) epidemics on a scale-free network. At every discrete time step $t$, node $i$ is either susceptible $(S_i)$, infected $(I_i)$, or recovered $(R_i)$ with an imperfect immunity. $S_i$, $I_i$, and $R_i$ are mutually exclusive binary state variables, indicating the epidemiological state of node $i$ at time $t$. The number of neighboring nodes of node $i$ that are infected at time $t$ is given as $\sum_j a_{ij} I_j (t)$, where $a_{ij} \in \{0,1\}$ represents the presence $(a_{ij}=1)$ or absence $(a_{ij} = 0)$ of a link between nodes $i$ and $j$. All nodes are assumed to be born fully susceptible to the infection, and infection can only be transmitted from an infected neighbor at the rate $\beta \sum_{j=1}^N a_{ij}I_j$. Nodes with higher degrees are more likely to contract the infection simply because they have more neighbors, and naturally, more infected neighbors~\cite{PastorSatoras-Vespignani-1, SHIRLEY-Rushton2005}. Infected nodes die from the infection at rate $\nu$ or recover from the infection at rate $\alpha$. Recovered nodes obtain immunity upon recovery. This immunity gained is assumed to be imperfect and wanes at rate $\gamma$, after which the node becomes susceptible to infection again. 

In addition to epidemic dynamics, the population is subject to birth and death processes. Every node, regardless of its epidemic state, dies naturally at rate $\mu$. All nodes that die, whether natural or infection-induced, are removed from the population together with all their links. On the other hand, new nodes are added at a rate proportional to the population, $bN(t)$. These new nodes are attached to $m$ already existing nodes with a probability proportional to their degree $k$, according to the preferential attachment rule. As a result of the demographic processes, the size of the network may grow, shrink, or remain constant. The co-evolution of the epidemics and the network topology is illustrated in Fig.~\ref{fig:schematic}. 

We begin our simulation by first generating a scale-free network using the Barab\'{a}si-Albert model~\cite{Barabasi-Albert} with an initial $N(t=0)$ nodes. At the beginning of the simulation, a small number of initially infected nodes $I_0 \ll N(t=0)$ are chosen uniformly at random. The epidemic dynamics are simulated using a Monte Carlo framework with $M$ independent realizations for each pair of infection fatality rate $\nu$ and immunity waning rate $\gamma$. We keep the other parameters of the model fixed. For the time-series analysis of the epidemics, ensemble averages are computed at each time step by averaging the state variables across all realizations. To characterize the time evolution of the degree distribution, the node degrees from all realizations are aggregated (chained) into a single ensemble at each time step, effectively constructing a composite network of size $\sum_{i=1}^M N(t)$. This ensemble-based approach provides a statistically robust estimate of the instantaneous degree distribution, enabling us to track its temporal evolution. Unless stated otherwise, the following parameter values are used in this work: $b = 0.0005$, $\mu = 0.0001$, $N(t=0) = 10,000$, $\beta = 0.06$, $\alpha = 0.04$, $m = 4$, $I_0=10$, $M=100$.

\section{Results}

\begin{figure*}
  \centering
  \includegraphics[width=0.95\textwidth]{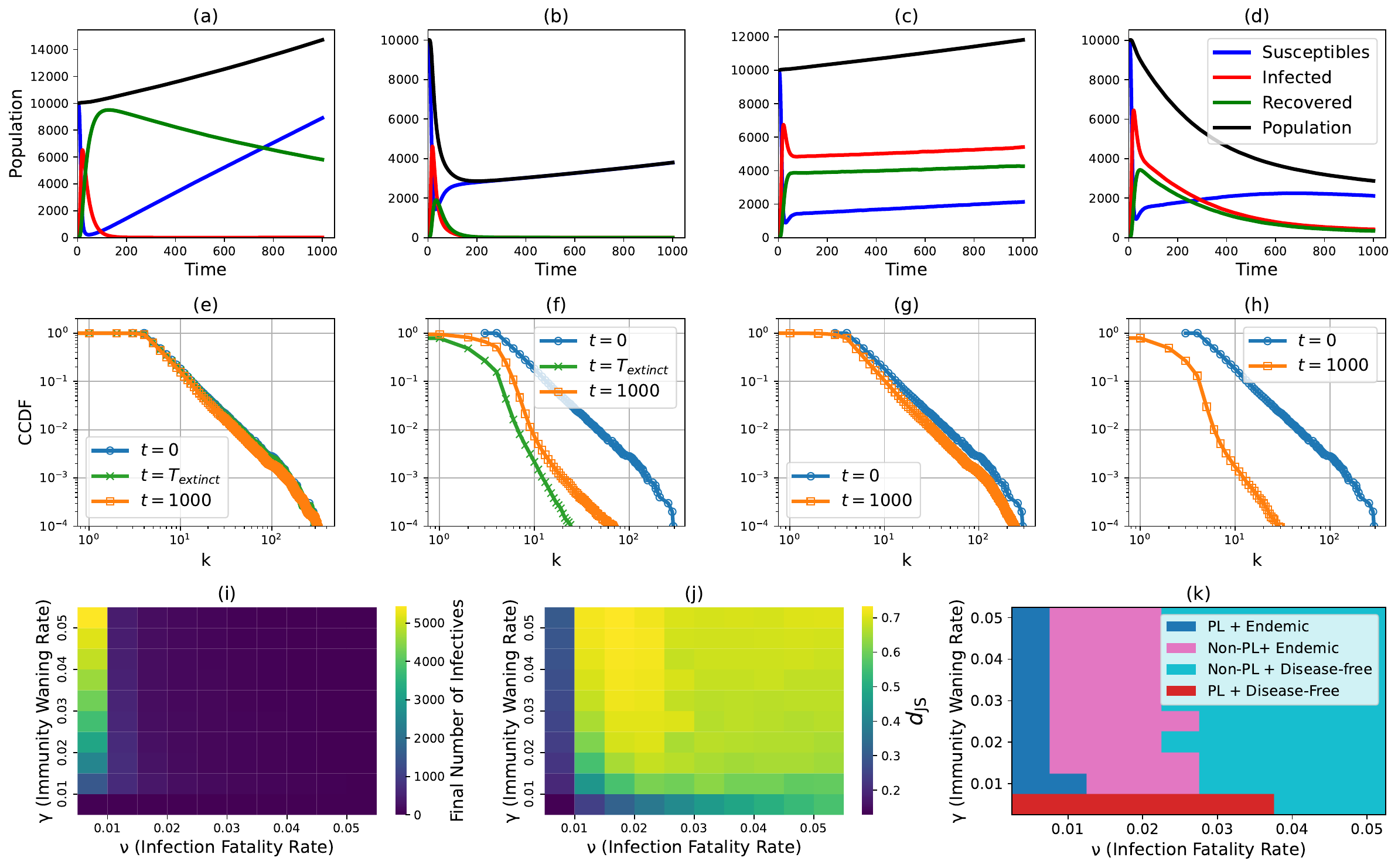}
  \caption{Summary of the results of our agent-based simulation. (a)--(d): The total population and the numbers of susceptible, infected, and recovered individuals as a function of time. (e)--(h): The complementary cumulative distribution function $(\mathrm{CCDF})$ of the degree distribution before the simulation $(t=0)$, when the epidemic goes extinct $(t=T_{\text{extinct}})$ and at the end of the simulation $(t=1000)$. (i) Heatmap of the final number of infected individuals for the parameter space, highlighting the existence of the disease-free and endemic regions. (j) Heatmap of the Jensen-Shannon distance between the degree distributions at $t=0$ and at $(t=T_{\text{extinct}})$ for the disease-free cases or $t=1000$ for the endemic cases, highlighting the transition of the degree distribution from the initial power-law distribution to an exponential distribution at both extremes. (k) Here, we classify regions where $d_{\mathrm{JS}}  \le 0.4$ as the power-law region, and regions with $d_{\mathrm{JS}} > 0.4$ as the non-power-law region. The sub-plot shows the phase diagram for the epidemic-topology coevolution, showing four distinct regions. These regions are highlighted in; (a), (e): The power-law and disease-free region, $\nu = 0.0005, \gamma = 0.0005$. (b), (f): The non-power-law and disease free region (Non-PL + Disease-free), $\nu = 0.05, \gamma = 0.05$. (c), (g): The power-law and endemic region (PL + Endemic), $\nu = 0.0005, \gamma = 0.045$. (d),(h): The non-power-law and endemic region (Non-PL + Endemic), $\nu = 0.015, \gamma = 0.035$.}
  \label{fig:results}
\end{figure*}

We are interested in two dynamical processes occurring simultaneously in the network. The first is the epidemic dynamics. Our results show the existence of two distinct epidemic regions: a disease-free region, in which infection eventually dies out (Fig.~\ref{fig:results}(a, b)), and an endemic region, in which infection remains in the population (Fig.~\ref{fig:results}(c, d)). Fig.~\ref{fig:results}(i) shows the dependence of the final number of infected nodes on the immunity waning and infection fatality rates.

The disease becomes endemic as a result of an increased immunity waning rate coupled with a decreased removal of nodes via infection death. Here, a key observation is that high-degree nodes have a greater probability of being infected. When disease fatality is low, these hub nodes remain longer in the population, act as super spreaders, and contribute to the sustained infection. When this super-spreading is coupled with decreasing efficacy of the immunity (increasing waning rates), recovered nodes quickly lose their immunity and become susceptible again, thereby increasing the endemicity of the disease. In other words, the epidemic dynamics effectively reduces to an SIS model in this region.

Conversely, in the disease-free region, the epidemic is eliminated through either increased infectious death rates or high immunity levels. An increased infectious death rate ensures that infected nodes are quickly removed from the population. At higher infectious death rates, the rate at which the infected nodes are removed is greater than the rate at which they transmit the infection to others. Hence, the infection dies out much faster and does not persist in the population for a long time. Lower waning rates, on the other hand, correspond to a near-perfect immunity, making recovered agents (especially the hubs) remain immune for a longer period. This sustained immunity helps to break the spread of the epidemics in the population, resulting in a disease-induced herd immunity~\cite{Britton2020, Takayuki-Zhara-Abbas-Mikko-Jari}.

The other dynamical process of interest is the evolution of the network. The topology of the network evolves via the addition and preferential attachment of new nodes through birth, coupled with the removal of nodes that die either naturally or from infection. We capture the degree distribution at three different time points: at the beginning of the simulation of the epidemic ($t=0$), when the epidemic dies out ($t=T_{\text{extinct}}$), and at the end of the simulation ($t=1000$). We observe that for some parameters, the initial power-law degree distribution is preserved at the end of the simulation, as shown in Fig.~\ref{fig:results}(e, g). On the other hand, there is a deviation from the initial power-law degree distribution for some other parameter sets as shown in Fig.~\ref{fig:results}(f, h). 

To quantify these changes in the degree distribution, we use the Jensen-Shannon distance. Given two discrete probability distributions $P$ and $Q$, the Jensen-Shannon divergence~\cite{lin1991JSDivergence} is given as:
\begin{equation}
    \mathrm{JSD}(P,Q) = \frac{1}{2} \mathrm{KL}(P||M) + \frac{1}{2} \mathrm{KL}(Q||M),
    \label{JSD}
\end{equation}
where $$M = \frac{P+Q}{2},$$ and $$\mathrm{KL}(P||Q)= \sum_i P(i)\log \bigg(\frac{P(i)}{Q(i)} \bigg),$$ is the Kullback-Leibler divergence or relative entropy~\cite{Kullback-Leibler}. The Jensen-Shannon distance~\cite{endres2003JSDistance} is the square root of the Jensen-Shannon divergence, given by:
\begin{equation}
     d_{\mathrm{JS}}(P,Q) = \sqrt{\mathrm{JSD}(P,Q)},
\end{equation}
where $0 \le d_{\mathrm{JS}}(P,Q) \le 1$. $d_{\mathrm{JS}}$ close to zero indicates that both distributions are identical, while $d_{\mathrm{JS}}$ close to one shows that the distributions are maximally different.

Fig.~\ref{fig:results}(j) shows the Jensen-Shannon distance between the degree distributions at the start of the simulation and when the epidemic goes extinct for disease-free cases. For endemic cases, it shows the distance between the degree distributions at the start and end of the simulation. We observe a topological transition from a parameter region where the initial power-law degree distribution was preserved to a region where the power-law distribution was truncated, taking a non-power-law form.

The non-power-law region emerges because high infection fatality selectively removes highly connected nodes and their links, and degrades the scale-free topology of the network. When the infectious death rate is high, provided the waning rate is sufficiently large, hub nodes will eventually die due to continuous reinfection and are removed from the network. The removal of hub nodes progressively truncates the heavy-tailed degree distribution. As shown in Fig.~\ref{fig:results}(h), the initial network has a highest degree of 300, while the highest degree at the end of the simulation is about 30. This shows a depletion of the initial scale-free structure, leading to the emergence of the non-power-law region, as seen in Fig.~\ref{fig:results}(j).

In contrast, the power-law regime is preserved at low infection fatality or immunity waning rates. When disease fatality is low, the net growth process is dominated by birth and preferential attachment. This results in the preservation of the scale-free structure. The scale-free structure is also maintained at stronger immunity levels (low immunity waning rates). This holds even at a higher infection fatality rate, albeit with the loss of some of the hub nodes that were initially infected. This is because, if the initially infected hub nodes escape death, they do not easily get reinfected (since the immunity obtained from recovery is high), thereby reducing their chances of dying. These nodes continue to attract new neighbors via preferential attachment, thereby sustaining the scale-free structure of the network.

So far, we have examined the epidemic dynamics and the evolution of the network separately. We now consider how the epidemic dynamics influences the topological evolution of the scale-free network, and how the topology of the network simultaneously influences the dynamics of the epidemics. As a combination of the epidemic transition and the topological evolution of the network, we obtain four distinct regions as shown in the phase diagram of Fig.~\ref{fig:results}(k): (i) a disease-free region with the power-law degree distribution preserved, (ii) an endemic region with the power-law degree distribution preserved, (iii) a disease-free region with a non-power law degree distribution, and (iv) an endemic region with the degree distribution deviating from the initial power-law distribution. The emergence of these four regions demonstrates the bidirectional coupling between epidemic dynamics and network evolution. 

\begin{figure*}[ht!]
  \centering
  \includegraphics[width=1.0\textwidth]{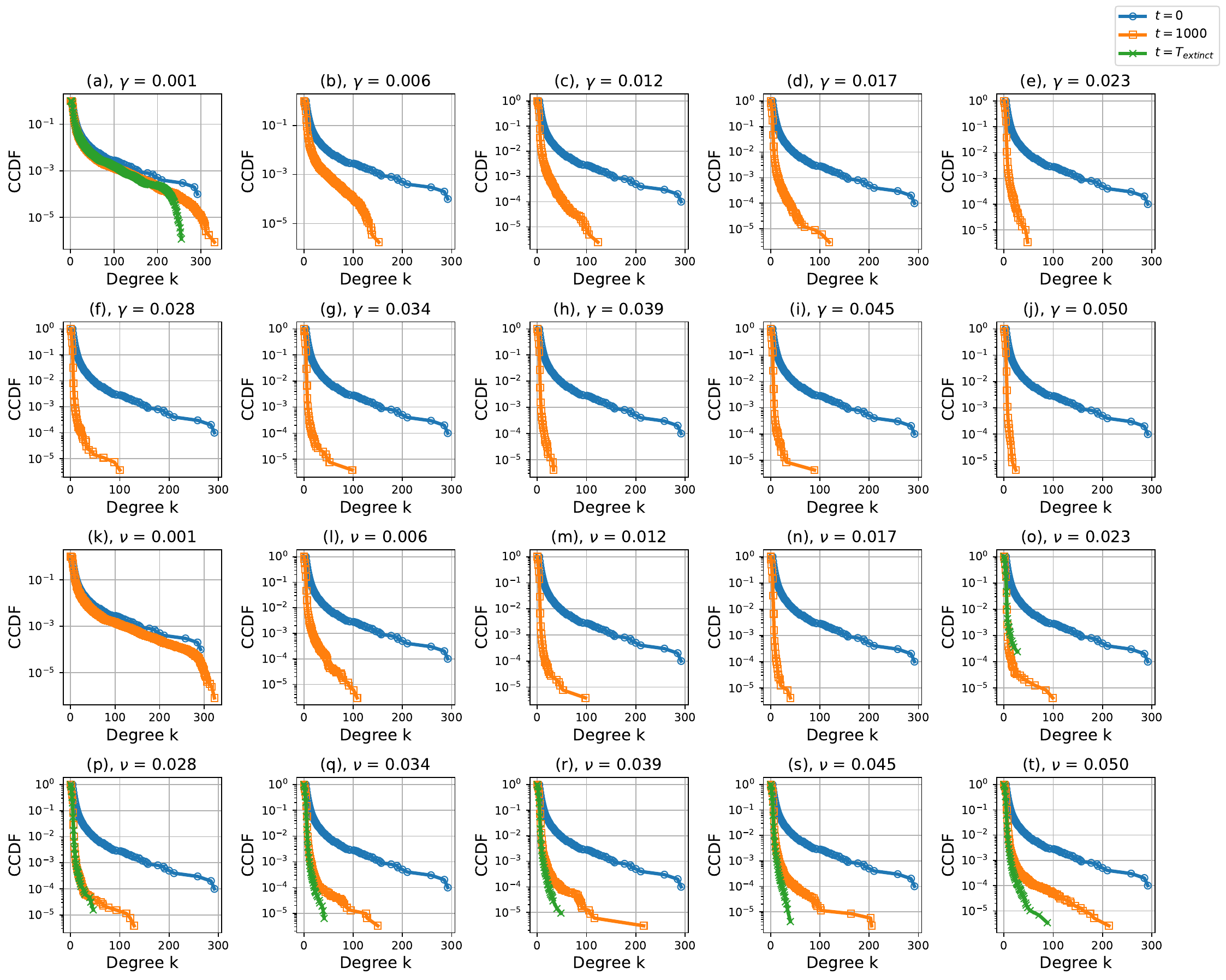}
  \caption{Complementary cumulative distribution function $(\mathrm{CCDF})$ of the degree distribution before the simulation $(t=0)$, when the epidemic goes extinct $(t=T_{\text{extinct}})$, and at the end of the simulation $(t=1000)$. The subplots (a)--(j) represent how the degree distribution evolves as the immunity waning rate increases. Here, the infection fatality rate is fixed at $\nu = 0.0115$, while the immunity waning rate ranges from $\gamma = 0.0001$ in (a), to $\gamma = 0.05$ in (j). On the other hand, the subplots (k)--(t) represent the evolution of the degree distribution, as the infection fatality rate increases. Here, the immunity waning rate is fixed at $\gamma = 0.0335$, while the infection fatality rate ranges from $\nu = 0.0001$ in (k) to $\nu = 0.05$ in (t). The appearance of the plots for $(t=T_{\text{extinct}})$ indicates that the number of infected individuals goes to zero in at least one of the $M$ independent realizations.}
  \label{fig:dependence}
\end{figure*}

The coevolution between epidemic dynamics and network topology is governed by the interplay between waning immunity and infection-induced removal. The epidemic process influences the topological evolution of the network through the selective removal of infected nodes and their associated links, which acts as a targeted attack on the scale-free structure. When infection-induced removal is sufficiently small, preferential attachment dominates the network evolution, preserving the scale-free degree distribution. In this regime, increasing the immunity waning rate primarily affects the epidemic dynamics, driving a transition from a disease-free state to an endemic state while the underlying power-law topology remains intact (Fig.~\ref{fig:results}(k)). In contrast, at intermediate infection-induced removal rates, the epidemic not only alters its own persistence but also significantly reshapes the network structure through the progressive loss of highly connected nodes. This results in a concurrence of an epidemic transition from the disease-free to the endemic phase and a topological transition whereby the degree distribution deviates from a power-law. Namely, the degree distribution of the remaining subnetwork approaches an exponential form (Fig.~\ref{fig:results}(k), Fig.~\ref{fig:dependence}(a--j)). This observation is consistent with the findings of Moore et al.~\cite{Moore-Ghoshal-Newman} and Bauke et al.~\cite{Bauke-Moore-et.al2011}, who showed that sufficiently strong removal processes cause shrinking networks to lose their scale-free structure and approach exponential degree distributions.

The evolving topology simultaneously influences the epidemic dynamics by modifying the heterogeneity of the contact network and the availability of highly connected transmission pathways. As hubs are removed and the degree distribution departs from the power-law form, disease transmission becomes less efficient, reducing the ability of the epidemic to persist. Consequently, increasing the infection-induced removal rate drives a transition from the endemic to the disease-free phase, accompanied by a topological transition from power-law to non-power-law networks. At sufficiently large removal rates, infected nodes are removed rapidly, causing epidemic extinction and a temporary collapse of the scale-free structure. Following extinction, nodes are no longer removed by infection, and network growth once again dominates the dynamics. As a result, the population recovers and the degree distribution gradually returns to its initial power-law form (Fig.~\ref{fig:results}(b, f), Fig.~\ref{fig:dependence}(p--t)). These results demonstrate that epidemic dynamics and topological evolution are not independent processes, but instead form a coupled feedback system in which epidemic spread reshapes the network while the evolving network structure simultaneously regulates epidemic persistence and extinction.

\section{Analysis}

\begin{figure*}
  \centering
    \includegraphics[width=0.95\textwidth]{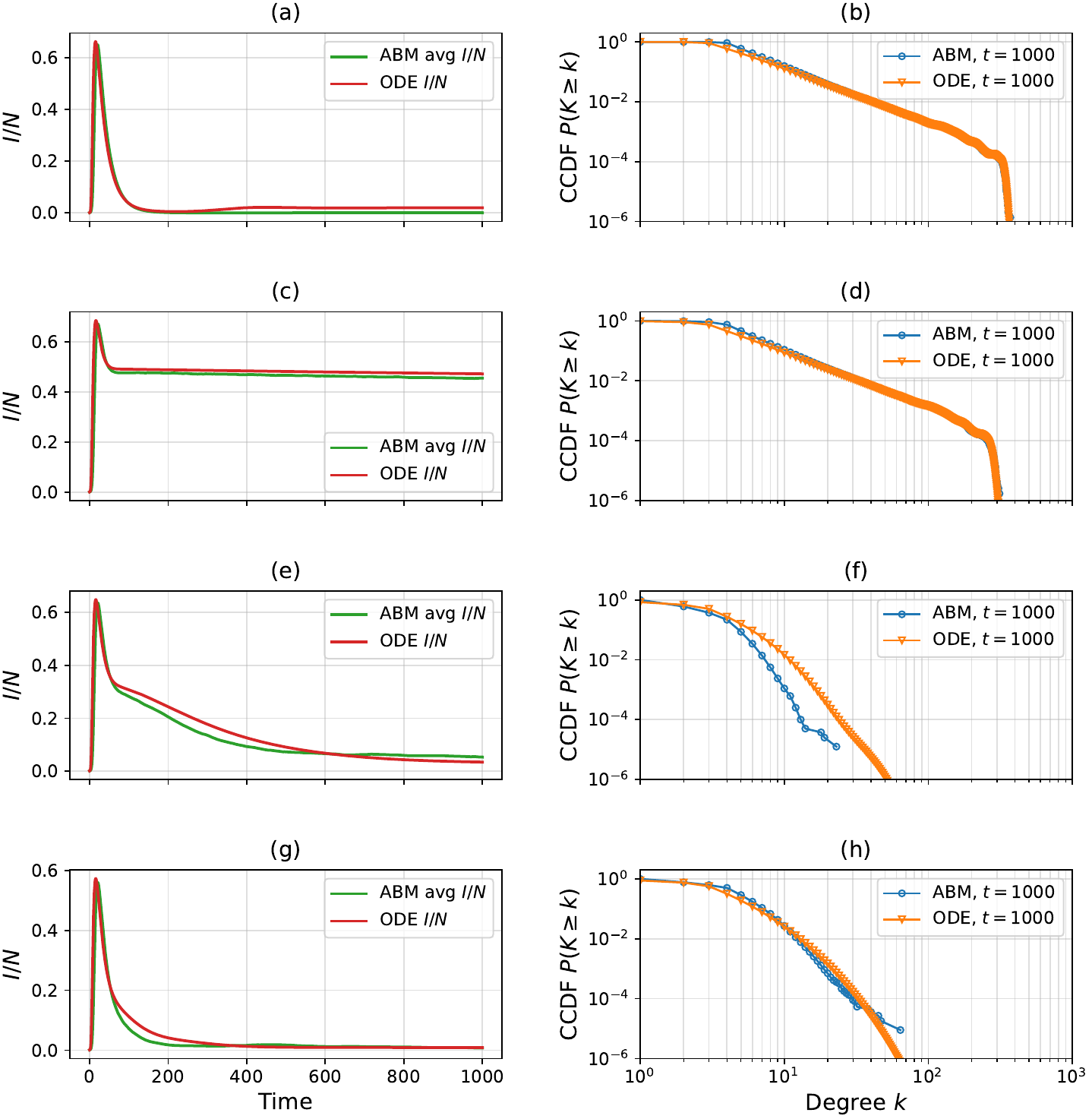}
  \caption{Comparison of the agent-based model (ABM) to the ODE model. The plots on the LHS show the proportion of infected as a function of time, for the ODE and the ABM, while the RHS compares the CCDF of the degree distribution between the ODE and the ABM, at the end of the simulation. (a): The power-law and disease free region, $\nu = 0.0005, \gamma = 0.0005.$ (b): The non-power-law and disease-free region, $\nu = 0.05, \gamma = 0.05.$ (c): The power-law and endemic region, $\nu = 0.0005, \gamma = 0.045.$ (d): The non power-law endemic region, $\nu = 0.015, \gamma = 0.035.$ }
  \label{fig:ABMvsODE}
\end{figure*}

 We complement the agent-based simulations with a degree-structured ordinary differential equation (ODE) framework that describes the macroscopic evolution of both the epidemic states and the network topology. In contrast to the stochastic agent-based model (ABM), the ODE formulation provides a deterministic description of the coupled epidemic-network dynamics. This enables us to identify the mechanisms underlying the observed epidemic and topological transitions. Our formulation is based on the heterogeneous mean-field approximation, which assumes statistical equivalence among nodes within the same degree class~\cite{PastorSatoras-et.al-ReviewPaper, Barrat-Barthelemy-Vespignani_Book}. Under this approximation, and assuming the absence of degree correlations, we derive the following degree-structured equations as an approximation to the ABM:

\begin{widetext}
    \begin{equation}
\begin{aligned}
    \frac{d S_k(t)}{d t} =\ & 
    b N(t) \delta_{k, m} - \beta k\Theta(t) S_k(t) + \gamma R_k(t) - \mu S_k(t)
    + b N(t) m \Pi_{k-1}(t) S_{k-1}(t) - b N (t)m \Pi_k(t) S_k(t) \\[5pt]
    &- k \big(\mu + \nu \Theta(t) \big) S_k(t) + (k+1) \big(\mu + \nu \Theta(t) \big) S_{k+1}(t),\\[5pt]
    \frac{d I_k(t)}{d t} =\ & 
    \beta k \Theta(t) S_k(t) - (\alpha + \nu) I_k(t) - \mu I_k(t) + b N(t) m \Pi_{k-1}(t) I_{k-1}(t) - b N(t) m \Pi_k(t) I_k(t) \\[5pt]
    &- k \big(\mu + \nu \Theta(t) \big) I_k(t) + (k+1) \big(\mu + \nu \Theta(t) \big) I_{k+1}(t),\\[5pt]
    \frac{d R_k(t)}{d t} =\ & 
    \alpha I_k(t) - \gamma R_k(t) - \mu R_k(t) + b N(t) m \Pi_{k-1}(t) R_{k-1}(t) - b N(t) m \Pi_k(t) R_k(t) \\[5pt]
    &- k\big(\mu + \nu \Theta(t) \big) R_k(t) + (k+1)\big(\mu + \nu \Theta(t) \big) R_{k+1}(t),
\end{aligned}
\label{ODE_General}
\end{equation}
\end{widetext}
where $\Pi_k(t)=k/[N(t)\langle k\rangle_t]$ is the probability that a newly added edge attaches to a specific node of degree $k$. The quantity $\Theta(t)$ denotes the probability that a randomly selected edge is connected to an infected node. Thus, a susceptible node of degree $k$ has, on average, $k\Theta(t)$ infected neighbors. If $I_k(t)$ denotes the number of infected nodes of degree $k$, then
\begin{equation}
    \Theta(t)
    =
    \frac{\sum_k k I_k(t)}
    {\sum_k k N_k(t)}
    =
    \frac{\sum_k k I_k(t)}
    {N(t)\langle k\rangle_t}.
    \label{theta}
\end{equation}
${\left<k\right>_t} = \sum_{k} kP_k(t)$ is the mean degree at time $t$, and $P_k(t)$ is the degree distribution of the network at time $t$. Here, the probability of a link pointing to an infected node is independent of the connectivity of the node the link emanates from. This is a simplifying modeling assumption, which may not necessarily hold in reality. Notice that, in the above formulation, the expected number of susceptible neighbors to which an infected node transmits the disease scales linearly with the degree $k$, so that highly connected nodes contribute more significantly to disease spread. However, the scaling can be modeled as constant, probabilistic, or nonlinear; see~\cite{LI-Chi-Yang2014} and the references therein.

\begin{figure*}[ht!]
  \centering
  \includegraphics[width=1.0\textwidth]{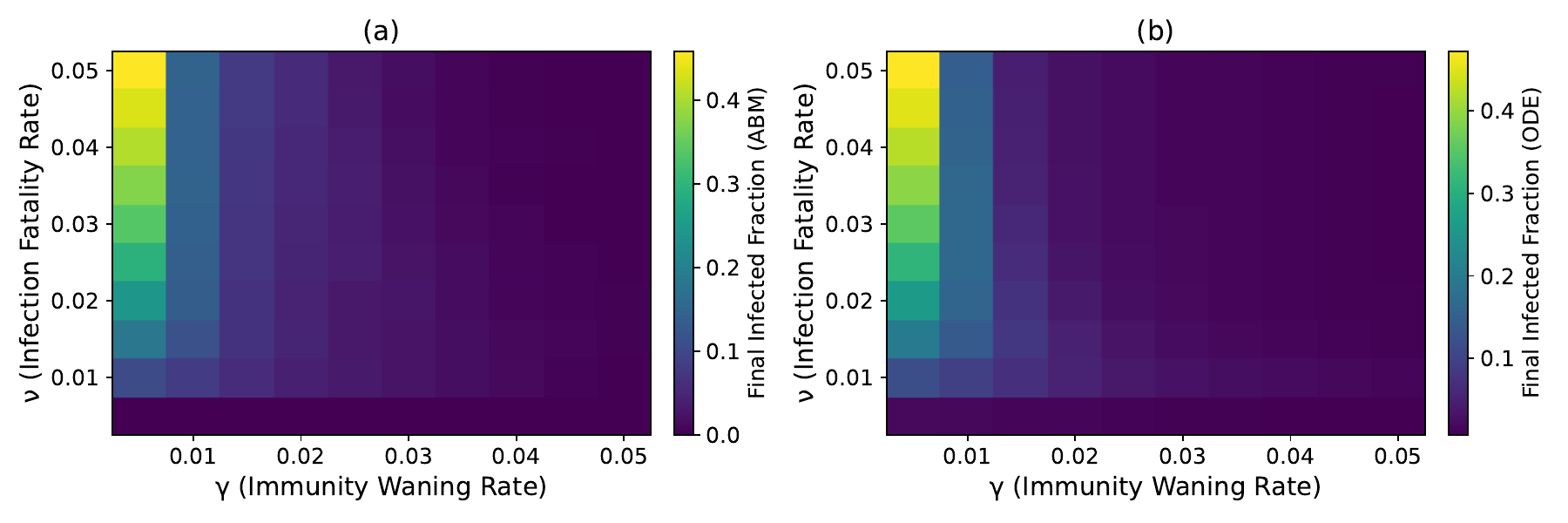}
  \caption{Final average fraction of infected nodes for (a) the ABM and (b) the ODE.}
  \label{fig:ABMvsODEDifference}
\end{figure*}

The above system correctly mirrors our ABM. New nodes of degree $m$ are added to the susceptible class per unit time, at a per capita birth rate $b$. The total number of new nodes added at each time $t$ is given as $bN(t) \delta_{k,m}$. These nodes are preferentially attached to already existing $X_k \in \{ S_k, I_k, R_k \} $ nodes in the network. When a node of degree $k-1$ gains an edge, it becomes a node of degree $k$. The total number of degree $k-1$ nodes that gain additional edges at time $t$ is given by $bN(t)m \Pi_{k-1} X_{k-1}(t)$. Similarly, nodes of degree $k$ could gain an additional edge, and would no longer be of degree $k$. The total number of degree $k$ nodes that gain an additional edge at time $t$ is given as $bN(t)m \Pi_k X_k(t)$. 

The epidemic transition also follows a similar mechanism as in the ABM. Nodes are infected at rate $\beta k\Theta(t) S_k$, where $\Theta(t)$ is the heterogeneous mean field parameter that represents the probability that a neighbor of a node is infected at time $t$. A degree $k$ node has, on average, $k\Theta(t)$ infected neighbors; the probability that it loses an edge due to the infection-induced death of its neighbor is given by $k \nu \Theta(t)$. Therefore, the total number of degree $k$ nodes that would lose an edge due to the infectious death of a neighbor at time $t$ is given by $k \nu \Theta X_k(t)$. By a similar argument, the total number of degree $k+1$ nodes that lose an edge due to the infectious death of a neighbor at time $t$ is given by $(k+1) \nu \Theta X_{k+1}(t)$. Nodes also die naturally at the rate $\mu X_k(t)$. The total number of degrees $k$ and $k+1$ nodes that lose an edge due to the natural death of a neighbor is given as $k\mu X_k(t)$ and $(k+1) \mu X_{k+1}(t)$, respectively. Processes in which a node gains or loses more than one edge at each time step are neglected.

Fig.~\ref{fig:ABMvsODE} shows that the heterogeneous mean-field ODE well approximates our agent-based model (ABM), accurately reproducing both the epidemic dynamics and the associated network evolution across all regimes. In the power-law regimes (Fig.~\ref{fig:ABMvsODE}(a--d)), the ODE closely matches the temporal evolution of the infected fraction, capturing both the transient outbreak and the long-term behavior, whether disease-free or endemic. It also preserves the heavy-tailed degree distributions, consistent with the preferential-attachment growth mechanism. In the non–power-law regimes (Fig.~\ref{fig:ABMvsODE}(e--h)), the ODE successfully captures the transition toward truncated degree distributions driven by infection-induced node removal, as well as the corresponding changes in epidemic prevalence. These results are in strong agreement with our results in Fig.~\ref{fig:results}(i, j). Fig.~\ref{fig:ABMvsODEDifference} shows that the epidemic transition observed in ABM is indeed replicated by the ODE model. Overall, the ODE reproduces the phase structure and macroscopic behavior of the ABM across all parameter regions. 

Despite this strong agreement, systematic differences arise due to the deterministic nature of the ODE and the stochasticity inherent in the ABM. In the disease-free regimes, the ABM exhibits finite-time extinction, whereas the ODE approaches zero asymptotically and retains a small residual infected fraction. Similar behavior has been widely documented in epidemic modeling, where deterministic frameworks capture average dynamics but fail to reproduce extinction as a finite-time event, which instead arises from stochastic fluctuations in finite populations~\cite{keeling2008HumansAndAnimals, allen2008stochastic}. These discrepancies highlight the limitations of mean-field approximations in capturing finite-size effects and stochastic extinction, even as they qualitatively describe the average system behavior.

\section{Discussion and Conclusions}

We studied the coevolution of epidemic dynamics and network topology in a growing scale-free network with waning immunity and infection-induced mortality. Using an agent-based SIRS model and a heterogeneous mean-field approximation, we found that epidemic dynamics and network structure are coupled through a feedback mechanism in which infection-induced node removal reshapes the degree distribution, while the evolving topology simultaneously regulates disease transmission and persistence. This feedback gives rise to distinct epidemic and topological regimes characterized by transitions between endemic and disease-free states, as well as between power-law and non-power-law degree distributions. Our results demonstrate that the long-term behavior of both the epidemic and the network emerges from the interplay between disease spread, node removal, and network growth, rather than from any of these processes acting independently.

We have shown that when the immunity derived is sufficiently strong, the epidemic generates a transient outbreak before eventually dying out, largely independent of the infection-induced mortality rate. In this regime, the dynamics approach those of a classical SIR model, since the waning rate is sufficiently small that recovered individuals remain protected for long periods of time. Such behavior is consistent with diseases that induce durable immune memory, including smallpox, measles, mumps, and rubella, for which immunity can persist for decades following infection or vaccination\mbox{~\cite{Hammarlund2003-Smallpox, Davidkin2008-MMR, SLIFKA1996}}. In contrast, when the infection-induced removal rate is sufficiently small, infected individuals remain in the population for longer periods, allowing sustained transmission and the emergence of an endemic state. In this regime, increasing the waning rate replenishes the susceptible population and further promotes disease persistence, consistent with classical epidemic theory~\cite{AndersonMay1991,keeling2008HumansAndAnimals}. At the same time, the low removal rate allows network growth through preferential attachment to dominate the dynamics, preserving the scale-free degree distribution. Similar preservation of scale-free structure under weak removal processes was reported by Moore et al.~\cite{Moore-Ghoshal-Newman}.

Although highly simplified, our model provides a possible mechanistic explanation for how epidemic outbreaks can temporarily reshape social contact networks. In the present framework, infection-induced removal may be interpreted more broadly as any process that removes infectious individuals from the transmission network, including death, quarantine, isolation, or hospitalization. When the removal rate is sufficiently large, highly connected nodes are preferentially eliminated from the network, reducing transmission opportunities and eventually driving the epidemic to extinction. This mechanism is qualitatively consistent with observations from outbreaks such as SARS in 2002--2003 and Ebola outbreaks in sub-Saharan Africa (Congo 1995, Uganda 2000, West Africa 2014--2016), where aggressive isolation and quarantine measures, together with high case fatality rates, contributed to the rapid suppression of transmission chains~\cite{Lipsitch2003,Riley2003,Chowell2004,Frieden2015Ebola}. In network terms, these interventions effectively removed infectious individuals from the contact structure and reduced the connectivity available for sustained disease spread. Following epidemic extinction, social interactions and mobility patterns gradually recovered, allowing contact networks to re-establish many of their pre-epidemic structural characteristics. Our results suggest that such recovery may be viewed as the re-emergence of network heterogeneity once the epidemic-driven removal pressure is relaxed.

Although our model predicts that the network can gradually recover its scale-free structure after epidemic extinction, real-world social systems may not always return to their pre-epidemic state. Major epidemics often induce behavioral, institutional, and demographic changes that persist long after transmission has ceased. Historical examples such as the Black Death have been associated with profound and long-lasting changes in social and economic organization~\cite{Herlihy1997BlackDeath}, while the COVID-19 pandemic accelerated the adoption of remote communication technologies and altered patterns of social interaction and human mobility behaviors~\cite{DeFilippis2022DigitalCommunication,Fujiwara-et.all2020MobilityTokyoCOVID19}. From the perspective of network science, such changes may be viewed as long-term modifications of the underlying contact structure. Although these processes are beyond the scope of the present model, our results suggest that epidemic-driven alterations of network topology may play an important role in shaping the long-term evolution of social systems.

In conclusion, our results demonstrate that epidemic dynamics and network topology cannot always be treated as independent processes. Instead, they may coevolve through a feedback mechanism in which disease spread alters network structure, while the evolving structure in turn influences disease persistence and extinction. This interplay can produce both epidemic and topological phase transitions.
More broadly, our findings underscore the importance of accounting for network evolution when studying epidemic processes on complex systems. Extending this framework to explore how adaptive behavior, intervention measures, and analytical approaches further shape the coevolutionary dynamics would be an important goal for future research. 

\begin{acknowledgments}
T.I.S.I. acknowledges support from JST SPRING, Grant Number JPMJSP2114. T.H. is supported by the Strategic Research Council (SRC) established within the Research Council of Finland (Decision Numbers: 364386 and 364371). N.F. is supported by JST PRESTO Grant Number JPMJPR21RA, Japan.
\end{acknowledgments}

\bibliography{Biblography}

@article{Barabasi-Albert,
author = {Albert-László Barabási  and Réka Albert },
title = {Emergence of Scaling in Random Networks},
journal = {Science},
volume = {286},
number = {5439},
pages = {509-512},
year = {1999},
doi = {10.1126/science.286.5439.509}}

@article{Barabasi-Jeong-Albert,
title = {Mean-field theory for scale-free random networks},
journal = {Physica A: Statistical Mechanics and its Applications},
volume = {272},
number = {1},
pages = {173-187},
year = {1999},
issn = {0378-4371},
doi = {https://doi.org/10.1016/S0378-4371(99)00291-5},
author = {Albert-László Barabási and Réka Albert and Hawoong Jeong}
}

@article{liljeros2001sexual,
  title={The web of human sexual contacts},
  author={Liljeros, Fredrik and Edling, Christofer R. and Amaral, L. A. Nunes and Stanley, H. Eugene and {\AA}berg, Yvonne},
  journal={Nature},
  volume={411},
  number={6840},
  pages={907--908},
  doi = {10.1038/35082140},
  year={2001},
  publisher={Nature Publishing Group}
}

@inproceedings{mislove2007measurement,
author = {Mislove, Alan and Marcon, Massimiliano and Gummadi, Krishna P. and Druschel, Peter and Bhattacharjee, Bobby},
title = {Measurement and analysis of online social networks},
year = {2007},
isbn = {9781595939081},
publisher = {Association for Computing Machinery},
address = {New York, NY, USA},
doi = {10.1145/1298306.1298311},
booktitle = {Proceedings of the 7th ACM SIGCOMM Conference on Internet Measurement},
pages = {29–42},
numpages = {14},
location = {San Diego, California, USA},
series = {IMC '07}
}

@article{guimera2005worldwide,
author = {R. Guimerà  and S. Mossa  and A. Turtschi  and L. A. N. Amaral },
title = {The worldwide air transportation network: Anomalous centrality, community structure, and cities' global roles},
journal = {Proceedings of the National Academy of Sciences},
volume = {102},
number = {22},
pages = {7794-7799},
year = {2005},
doi = {10.1073/pnas.0407994102}}

@article{PastorSatoras-Vespignani-1,
  title = {Epidemic Spreading in Scale-Free Networks},
  author = {Pastor-Satorras, Romualdo and Vespignani, Alessandro},
  journal = {Phys. Rev. Lett.},
  volume = {86},
  issue = {14},
  pages = {3200--3203},
  numpages = {0},
  year = {2001a},
  month = {Apr},
  publisher = {American Physical Society},
  doi = {10.1103/PhysRevLett.86.3200}
}

@article{PastorSatoras-Vespignani-2,
  title = {Epidemic dynamics and endemic states in complex networks},
  author = {Pastor-Satorras, Romualdo and Vespignani, Alessandro},
  journal = {Phys. Rev. E},
  volume = {63},
  issue = {6},
  pages = {066117},
  numpages = {8},
  year = {2001b},
  month = {May},
  publisher = {American Physical Society},
  doi = {10.1103/PhysRevE.63.066117}
}

@article{Newman2002,
  title = {Spread of epidemic disease on networks},
  author = {Newman, M. E. J.},
  journal = {Phys. Rev. E},
  volume = {66},
  issue = {1},
  pages = {016128},
  numpages = {11},
  year = {2002},
  month = {Jul},
  publisher = {American Physical Society},
  doi = {10.1103/PhysRevE.66.016128}
}

@article{Keeling-Eames2005,
  author       = {Keeling, Matt J. and Eames, Ken T. D.},
  title        = {Networks and epidemic models},
  journal      = {Journal of The Royal Society Interface},
  year         = {2005},
  volume       = {2},
  number       = {4},
  pages        = {295--307},
  doi          = {10.1098/rsif.2005.0051}
}

@article{PastorSatoras-et.al-ReviewPaper,
  title = {Epidemic processes in complex networks},
  author = {Pastor-Satorras, Romualdo and Castellano, Claudio and Van Mieghem, Piet and Vespignani, Alessandro},
  journal = {Rev. Mod. Phys.},
  volume = {87},
  issue = {3},
  pages = {925--979},
  numpages = {55},
  year = {2015},
  month = {Aug},
  publisher = {American Physical Society},
  doi = {10.1103/RevModPhys.87.925}
}

@article{Kermack-McKendrick1927,
  author    = {Kermack, W. O. and McKendrick, A. G.},
  title     = {A contribution to the mathematical theory of epidemics},
  journal   = {Proceedings of the Royal Society of London. Series A, Containing Papers of a Mathematical and Physical Character},
  year      = {1927},
  volume    = {115},
  number    = {772},
  pages     = {700--721},
  doi       = {10.1098/rspa.1927.0118}
}

@book{AndersonMay1991,
    author = {Anderson, Roy M and May, Robert M},
    title = {Infectious Diseases of Humans: Dynamics and Control},
    publisher = {Oxford University Press},
    year = {1991},
    month = {05},
    isbn = {9780198545996},
    doi = {10.1093/oso/9780198545996.001.0001},
}

@book{keeling2008HumansAndAnimals,
  title     = {Modeling Infectious Diseases in Humans and Animals},
  author    = {Keeling, Matt J. and Rohani, Pejman},
  year      = {2008},
  publisher = {Princeton University Press},
  address   = {Princeton, NJ},
  isbn      = {9780691116174}
}

@article{Hethcote,
  author    = {Hethcote, Herbert W.},
  title     = {The Mathematics of Infectious Diseases},
  journal   = {SIAM Review},
  year      = {2000},
  volume    = {42},
  number    = {4},
  pages     = {599--653},
  doi       = {10.1137/S0036144500371907}
}

@article{Diekmann1990,
  author    = {Diekmann, O. and Heesterbeek, J. A. P. and Metz, J. A. J.},
  title     = {On the definition and the computation of the basic reproduction ratio {$\mathcal{R}_0$} in models for infectious diseases in heterogeneous populations},
  journal   = {Journal of Mathematical Biology},
  year      = {1990},
  volume    = {28},
  number    = {4},
  pages     = {365--382},
  doi       = {10.1007/BF00178324}
}

@article{WattsStrogatz1998,
  author    = {Watts, Duncan J. and Strogatz, Steven H.},
  title     = {Collective dynamics of 'small-world' networks},
  journal   = {Nature},
  year      = {1998},
  volume    = {393},
  number    = {6684},
  pages     = {440--442},
  doi       = {10.1038/30918}
}

@article{Bansal2007,
  author    = {Bansal, Shweta and Grenfell, Bryan T. and Meyers, Lauren Ancel},
  title     = {When individual behaviour matters: homogeneous and network models in epidemiology},
  journal   = {Journal of The Royal Society Interface},
  year      = {2007},
  volume    = {4},
  number    = {16},
  pages     = {879--891},
  doi       = {10.1098/rsif.2007.1100},
  pmid      = {17640863},
  pmcid     = {PMC2394553}
}

@article{hethcote1973asymptotic,
  title={Asymptotic behavior in a deterministic epidemic model},
  author={Hethcote, Herbert W.},
  journal={Bulletin of Mathematical Biology},
  volume={35},
  number={5-6},
  pages={607--614},
  year={1973},
  month={November},
  doi={10.1007/BF02458365},
  pmid={4788626},
  publisher={Springer}
}

@article{rock2014dynamics,
  title={Dynamics of infectious diseases},
  author={Rock, Kat and Brand, Sam and Moir, Jo and Keeling, Matt J.},
  journal={Reports on Progress in Physics},
  volume={77},
  number={2},
  pages={026602},
  year={2014},
  publisher={IOP Publishing},
  doi={10.1088/0034-4885/77/2/026602}
}

@article{Xie-Ishfaq-ThankGod-Elza-Seno,
  title = {What Influence Could the Acceptance of Visitors Cause on the Epidemic Dynamics of a Reinfectious Disease?: A Mathematical Model},
  author = {Xie, Ying. and Ahmad, Ishfaq. and Ikpe, ThankGod I. S. and Sofia, Elza F. and Seno, Hiromi},
  journal = {Acta Biotheor},
  volume = {72},
  number = {3},
  pages = {3},
  year = {2024},
  doi = {10.1007/s10441-024-09478-w}
}

@article{DJD_Price1965,
author = {Derek J. de Solla Price },
title = {Networks of Scientific Papers},
journal = {Science},
volume = {149},
number = {3683},
pages = {510-515},
year = {1965},
doi = {10.1126/science.149.3683.510}
}

@article{Dorogovtsev-et.al,
  title = {Structure of Growing Networks with Preferential Linking},
  author = {Dorogovtsev, S. N. and Mendes, J. F. F. and Samukhin, A. N.},
  journal = {Phys. Rev. Lett.},
  volume = {85},
  issue = {21},
  pages = {4633--4636},
  numpages = {0},
  year = {2000},
  month = {Nov},
  publisher = {American Physical Society},
  doi = {10.1103/PhysRevLett.85.4633}
}

@article{Budnick-el.al_Isreal2022,
  title = {Structure of networks that evolve under a combination of growth and contraction},
  author = {Budnick, Barak and Biham, Ofer and Katzav, Eytan},
  journal = {Phys. Rev. E},
  volume = {106},
  issue = {4},
  pages = {044305},
  numpages = {15},
  year = {2022},
  month = {Oct},
  publisher = {American Physical Society},
  doi = {10.1103/PhysRevE.106.044305}
}

@article{Moore-Ghoshal-Newman,
  title = {Exact solutions for models of evolving networks with addition and deletion of nodes},
  author = {Moore, Cristopher and Ghoshal, Gourab and Newman, M. E. J.},
  journal = {Phys. Rev. E},
  volume = {74},
  issue = {3},
  pages = {036121},
  numpages = {8},
  year = {2006},
  month = {Sep},
  publisher = {American Physical Society},
  doi = {10.1103/PhysRevE.74.036121}
}

@article{Bauke-Moore-et.al2011,
  author    = {Bauke, Heiko and Moore, Cristopher and Rouquier, Jean-Baptiste and Sherrington, David and Rogers, Tim},
  title     = {Topological phase transition in a network model with preferential attachment and node removal},
  journal   = {The European Physical Journal B},
  year      = {2011},
  volume    = {83},
  number    = {4},
  pages     = {519--524},
  doi       = {10.1140/epjb/e2011-20346-0}
}

@article{Lee.M.J-et.al2022,
  title = {Degree distributions under general node removal: Power-law or Poisson?},
  author = {Lee, Mi Jin and Kim, Jung-Ho and Goh, Kwang-Il and Lee, Sang Hoon and Son, Seung-Woo and Lee, Deok-Sun},
  journal = {Phys. Rev. E},
  volume = {106},
  issue = {6},
  pages = {064309},
  numpages = {13},
  year = {2022},
  month = {Dec},
  publisher = {American Physical Society},
  doi = {10.1103/PhysRevE.106.064309}
}

@article{LI-Chi-Yang2014,
title = {Analysis of epidemic spreading of an SIRS model in complex heterogeneous networks},
journal = {Communications in Nonlinear Science and Numerical Simulation},
volume = {19},
number = {4},
pages = {1042-1054},
year = {2014},
issn = {1007-5704},
doi = {https://doi.org/10.1016/j.cnsns.2013.08.033},
author = {Chun-Hsien Li and Chiung-Chiou Tsai and Suh-Yuh Yang}
}

@article{HUANG-Chen-Chen2017,
title = {Global dynamics of a network-based SIQRS epidemic model with demographics and vaccination},
journal = {Communications in Nonlinear Science and Numerical Simulation},
volume = {43},
pages = {296-310},
year = {2017},
issn = {1007-5704},
doi = {https://doi.org/10.1016/j.cnsns.2016.07.014},
author = {Shouying Huang and Fengde Chen and Lijuan Chen}
}

@Article{Hu-et.al,
title = {Global dynamics of an SIRS model with demographics and transfer from infectious to susceptible on heterogeneous networks},
journal = {Mathematical Biosciences and Engineering},
volume = {16},
number = {5},
pages = {5729-5749},
year = {2019},
issn = {1551-0018},
doi = {10.3934/mbe.2019286},
author = {Haijun Hu and  Xupu Yuan and  Lihong Huang and  Chuangxia Huang},
}

@article{Demirel-Barter-Thilo,
  author    = {Demirel, G{\"o}khan and Barter, Edmund and Gross, Thilo},
  title     = {Dynamics of epidemic diseases on a growing adaptive network},
  journal   = {Scientific Reports},
  year      = {2017},
  volume    = {7},
  pages     = {42352},
  doi       = {10.1038/srep42352}
}

@article{SHIRLEY-Rushton2005,
title = {The impacts of network topology on disease spread},
journal = {Ecological Complexity},
volume = {2},
number = {3},
pages = {287-299},
year = {2005},
issn = {1476-945X},
doi = {https://doi.org/10.1016/j.ecocom.2005.04.005},
author = {Mark D.F. Shirley and Steve P. Rushton}
}

@article{Britton2020,
  author = {Britton, Tom and Ball, Frank and Trapman, Pieter},
  title = {A mathematical model reveals the influence of population heterogeneity on herd immunity to SARS-CoV-2},
  journal = {Science},
  year = {2020},
  volume = {369},
  number = {6505},
  pages = {846--849},
  doi = {10.1126/science.abc6810}
}

@article{
Takayuki-Zhara-Abbas-Mikko-Jari,
author = {Takayuki Hiraoka  and Zahra Ghadiri  and Abbas K. Rizi  and Mikko Kivelä  and Jari Saramäki },
title = {Strength and weakness of disease-induced herd immunity in networks},
journal = {Proceedings of the National Academy of Sciences},
volume = {122},
number = {28},
pages = {e2421460122},
year = {2025},
doi = {10.1073/pnas.2421460122}}

@article{lin1991JSDivergence,
  title={Divergence measures based on the Shannon entropy},
  author={Lin, Jianhua},
  journal={IEEE Transactions on Information Theory},
  volume={37},
  number={1},
  pages={145--151},
  year={1991},
  publisher={IEEE},
  doi={10.1109/18.61115}
}

@article{Kullback-Leibler,
author = {S. Kullback and R. A. Leibler},
title = {{On Information and Sufficiency}},
volume = {22},
journal = {The Annals of Mathematical Statistics},
number = {1},
publisher = {Institute of Mathematical Statistics},
pages = {79 -- 86},
year = {1951},
doi = {10.1214/aoms/1177729694}
}

@article{endres2003JSDistance,
  title={A new metric for probability distributions},
  author={Endres, Dominik Maria and Schindelin, Johannes E},
  journal={IEEE Transactions on Information Theory},
  volume={49},
  number={7},
  pages={1858--1860},
  year={2003},
  publisher={IEEE},
  doi={10.1109/TIT.2003.813506}
}

@inbook{Barrat-Barthelemy-Vespignani_Book,
  place={Cambridge},
  title={Epidemic spreading in population networks},
  booktitle={Dynamical Processes on Complex Networks},
  publisher={Cambridge University Press},
  author={Barrat, Alain and Barthélemy, Marc and Vespignani, Alessandro},
  year={2008},
  pages={180--215},
  doi={10.1017/CBO9780511791383.010}
}

@book{allen2008stochastic,
  title     = {An Introduction to Stochastic Processes with Applications to Biology},
  author    = {Allen, Linda J. S.},
  year      = {2008},
  publisher = {Chapman and Hall/CRC},
  address   = {Boca Raton, FL},
  isbn      = {9781584886419}
}

@article{Hammarlund2003-Smallpox,
  author    = {Erika Hammarlund and 
               Matthew W. Lewis and 
               Scott G. Hansen and 
               Lisa I. Strelow and 
               Jay A. Nelson and 
               Gary J. Sexton and 
               Jon M. Hanifin and 
               Mark K. Slifka},
  title     = {Duration of antiviral immunity after smallpox vaccination},
  journal   = {Nature Medicine},
  year      = {2003},
  month     = {Sep},
  volume    = {9},
  number    = {9},
  pages     = {1131--1137},
  issn      = {1546-170X},
  doi       = {10.1038/nm917}
}

@article{Davidkin2008-MMR,
  author    = {I. Davidkin and 
               S. Jokinen and 
               M. Broman and 
               P. Leinikki and 
               H. Peltola},
  title     = {Persistence of measles, mumps, and rubella antibodies in an {MMR}-vaccinated cohort: a 20-year follow-up},
  journal   = {The Journal of Infectious Diseases},
  year      = {2008},
  month     = {Apr},
  volume    = {197},
  number    = {7},
  pages     = {950--956},
  doi       = {10.1086/528993},
  pmid      = {18419470}
}

@article{SLIFKA1996,
title = {Long-term humoral immunity against viruses: revisiting the issue of plasma cell longevity},
journal = {Trends in Microbiology},
volume = {4},
number = {10},
pages = {394-400},
year = {1996},
issn = {0966-842X},
doi = {https://doi.org/10.1016/0966-842X(96)10059-7},
author = {Mark K. Slifka and Rafi Ahmed}
}

@article{Lipsitch2003,
  author    = {Lipsitch, Marc and 
               Cohen, Ted and 
               Cooper, Ben and 
               Robins, James M. and 
               Ma, Stefan and 
               James, Lyn and 
               Gopalakrishna, Gowri and 
               Chew, Suok Kai and 
               Tan, Chorh Chuan and 
               Samore, Matthew H. and 
               Fisman, David and 
               Murray, Megan},
  title     = {Transmission dynamics and control of severe acute respiratory syndrome},
  journal   = {Science},
  year      = {2003},
  volume    = {300},
  number    = {5627},
  pages     = {1966--1970},
  doi       = {10.1126/science.1086616}
}

@article{Riley2003,
  author    = {Riley, Steven and 
               Fraser, Christophe and 
               Donnelly, Christl A. and 
               Ghani, Azra C. and 
               Abu-Raddad, Laith J. and 
               Hedley, Anthony J. and 
               Leung, Gabriel M. and 
               Ho, Lai-Ming and 
               Lam, Tze-Wai and 
               Thach, Thuan Q. and 
               Chau, Patrick and 
               Chan, Kit Ping and 
               Lo, Siu-Van and 
               Leung, Pak-Yin and 
               Tsang, Thomas and 
               Ho, Wun-Yin and 
               Lee, Kwok-Hung and 
               Lau, Eric M. C. and 
               Ferguson, Neil M. and
               Anderson, Roy M.},
  title     = {Transmission dynamics of the etiological agent of SARS in Hong Kong: impact of public health interventions},
  journal   = {Science},
  year      = {2003},
  volume    = {300},
  number    = {5627},
  pages     = {1961--1966},
  doi       = {10.1126/science.1086478}
}

@article{Chowell2004,
  author    = {Chowell, Gerardo and 
               Hengartner, Nicolas W. and 
               Castillo-Chavez, Carlos and 
               Fenimore, Paul W. and 
               Hyman, James M.},
  title     = {The basic reproductive number of {Ebola} and the effects of public health measures: the cases of {Congo} and {Uganda}},
  journal   = {Journal of Theoretical Biology},
  year      = {2004},
  volume    = {229},
  number    = {1},
  pages     = {119--126},
  doi       = {10.1016/j.jtbi.2004.03.006}
}

@article{Frieden2015Ebola,
  author  = {Frieden, Thomas R. and Damon, Inger K.},
  title   = {Ebola in {West} {Africa}---{CDC}'s Role in Epidemic Detection, Control, and Prevention},
  journal = {Emerging Infectious Diseases},
  year    = {2015},
  volume  = {21},
  number  = {11},
  pages   = {1897--1905},
  month   = {Nov},
  doi     = {10.3201/eid2111.150949},
  pmid    = {26484940},
  pmcid   = {PMC4622264}
}

@book{Herlihy1997BlackDeath,
  title     = {The Black Death and the Transformation of the West},
  author    = {Herlihy, David},
  editor    = {Cohn, Jr., Samuel K.},
  year      = {1997},
  publisher = {Harvard University Press},
  address   = {Cambridge, MA},
  isbn      = {9780674076136}
}

@article{DeFilippis2022DigitalCommunication,
  title = {The impact of COVID-19 on digital communication patterns},
  author = {DeFilippis, Evan and Impink, Stephen Michael and Singell, Madison and Polzer, Jeffrey T. and Sadun, Raffaella},
  journal = {Humanities and Social Sciences Communications},
  volume = {9},
  number = {1},
  pages = {1--11},
  year = {2022},
  month = {May},
  day = {23},
  doi = {10.1057/s41599-022-01190-9}
}

@article{Fujiwara-et.all2020MobilityTokyoCOVID19,
  author    = {Yabe, Takahiro and Tsubouchi, Keisuke and Fujiwara, Naoya and Wada, Takaaki and Sekimoto, Yoshihide and Ukkusuri, Satish V.},
  title     = {Non-compulsory measures sufficiently reduced human mobility in Tokyo during the COVID-19 epidemic},
  journal   = {Scientific Reports},
  year      = {2020},
  volume    = {10},
  number    = {1},
  pages     = {18053},
  doi       = {10.1038/s41598-020-75033-5},
  publisher = {Nature Publishing Group}
}
\end{document}